%% file: main.tex
\documentclass[reprint,superscriptaddress,amsmath,amssymb,aps,prd]{revtex4-2}
\usepackage{graphicx}
\usepackage{dcolumn}
\usepackage{bm}
\usepackage[mathlines]{lineno}
\usepackage[hidelinks]{hyperref}
\usepackage{orcidlink}
\usepackage{misc/marco}


\begin{document}

\title{Constraints on Cosmic Strings from the Curl-Mode CMB Lensing Power Spectrum measured by ACT DR6}

\input{misc/authorslist}
\date{\today}

\input{sections/0abstract}

\maketitle

\input{sections/1introduction}
\input{sections/2theory}
\input{sections/3likelihood}
\input{sections/4discussion}

\input{sections/5conclusions}
\input{sections/6akg}
\input{sections/7data}


\bibliography{misc/ref}

\end{document}

%% file: misc/authorslist.tex
\author{A. I. Lonappan\,\orcidlink{0000-0003-1200-9179}}
\email{alonappan@ucsd.edu}
\affiliation{Department of Physics, University of California, San Diego, CA 92093, USA}

\author{K. Ramesh\,\orcidlink{0009-0007-1955-5670}}
\email{keerthanaramesht@gmail.com}
\affiliation{Department of Physics, NITK Surathkal, NH 66, Srinivasnagar Surathkal, Mangalore, Karnataka 575025}

\author{T. Namikawa\,\orcidlink{0000-0003-1200-9179}}
\affiliation{Center for Data-Driven Discovery, Kavli IPMU (WPI), UTIAS,
The University of Tokyo, Kashiwa, 277-8583, Japan}

\author{F. J. Qu\,\orcidlink{0000-0003-1200-9179}}
\affiliation{Kavli Institute for Particle Astrophysics and Cosmology,
Stanford University, 452 Lomita Mall, Stanford, CA, 94305, USA}
\affiliation{Department of Physics, Stanford University, 382 Via Pueblo Mall, Stanford, CA, 94305, USA}
\affiliation{SLAC National Accelerator Laboratory, 2575 Sand Hill Road, Menlo Park, California 94025, USA}

\author{B. Keating\,\orcidlink{0000-0003-3118-5514}}
\affiliation{Department of Physics, University of California, San Diego, CA 92093, USA}


%% file: sections/0abstract.tex
\begin{abstract}
A network of cosmic strings is one of the few well-motivated cosmological sources of vector and tensor metric perturbations on the largest observable scales. Such perturbations imprint a characteristic curl component in the deflection angle of cosmic microwave background (CMB) photons that, unlike the scalar lensing potential, vanishes for adiabatic density fluctuations at linear order. We exploit the curl-mode lensing reconstruction released as part of the Atacama Cosmology Telescope (ACT) Data Release~6 (DR6), based on five seasons of temperature and polarization data covering $9400~\mathrm{deg}^2$ of sky, to set new constraints on the dimensionless string tension $G\mu$ and the inter-commutation (reconnection) probability $P$. Modelling the string-induced curl power spectrum within the velocity-dependent one-scale framework, we obtain a $2\sigma$ upper bound on the combination $G\mu P^{-1}\le 3.5\times 10^{-5}$ in the small-$P$ regime, and $G\mu \le 5.0\times 10^{-5}$ at $2\sigma$ assuming the canonical Nambu-Goto value $P=1$. Combining the ACT DR6 curl bandpowers with the Planck 2013 curl-mode reconstruction, which extends down to $L_{\rm min}=2$, tightens these bounds to $G\mu P^{-1}\le 3.2\times 10^{-5}$ and $G\mu\le 4.3\times 10^{-5}$ ($2\sigma$). These represent the tightest constraints on cosmic strings derived from the curl-mode CMB lensing power spectrum to date. Using the ACT data alone, compared to the ACT 2008-season analysis, the ACT DR6 constraint on $G\mu P^{-1}$ is nearly an order of magnitude tighter.
\end{abstract}

%% file: sections/1introduction.tex
\section{Introduction}\label{sec:intro}

Weak gravitational lensing of the cosmic microwave background (CMB) has matured into a high-precision probe of the matter distribution in the late Universe. Following the first detections~\cite{Smith:2007rg,Hirata:2008cb,Das:2011ak,vanEngelen:2012va,Bleem:2012gm,Sherwin:2012mn,Das:2013zf,Ade_2014}, the lensing power spectrum is now measured at tens of $\sigma$ by current ground-based and space-borne experiments. Among current ground-based experiments, the Atacama Cosmology Telescope (ACT) has now released its Data Release~6 (DR6) lensing analysis~\cite{Qu:2023jia}, which exploits five seasons of temperature and polarization observations spanning $9400~\mathrm{deg}^2$ of sky and yields one of the highest signal-to-noise lensing power spectrum measurements to date from a single ground-based experiment, comparable in sensitivity to the Planck lensing reconstruction.

Although the dominant CMB lensing signal is generated by scalar metric perturbations along the line of sight, the deflection field $\bm{d}(\hat{\bm n})$ admits a clean parity decomposition into a gradient component $\phi$ and a divergence-free curl component $\Omega$~\cite{Kamionkowski:1996zd,Hirata:2003ka,Cooray:2005hm,Namikawa:2011cs},
\begin{equation}
\bm{d}(\hat{\bm n}) \;=\; \bm{\nabla}\phi(\hat{\bm n}) + (\star\bm{\nabla})\,\Omega(\hat{\bm n}),
\label{eq:deflection_decomp}
\end{equation}
where $\star$ denotes a counter-clockwise rotation by 90 degrees. At linear order in cosmological perturbation theory the curl component vanishes for scalar fluctuations, and is sourced exclusively by vector and tensor metric perturbations. While scalar perturbations do produce a curl component through second-order post-Born effects~\cite{Cooray:2002mj,Hirata:2003ka} and second-order perturbations from products of the first-order scalar modes \cite{Saga:2015:second-order-curl}, this signal is negligibly small on the angular scales relevant to the present analysis~\cite{Hirata:2003ka,Pratten:2016dsm,Robertson:2024qjl}.  The curl-mode lensing power spectrum, hereafter denoted by $\Cww$, is therefore a uniquely clean window onto non-scalar sources, and at the same time provides a powerful systematics test for any gradient-mode reconstruction~\cite{Qu:2023jia}.

A particularly compelling source of vector and tensor metric perturbations on cosmological scales is a network of cosmic strings. Such one-dimensional topological defects can form at symmetry-breaking phase transitions in the early Universe, and are also predicted to arise as fundamental F- and D-strings of cosmic-superstring origin at the end of brane inflation~\cite{Copeland:2003bj,Jackson:2004zg}. The phenomenology of the two classes is partly distinguished by the inter-commutation (reconnection) probability $P$: while ordinary field-theoretic strings always reconnect upon intersection ($P=1$), cosmic superstrings can have $P$ ranging over many decades below unity, depending on the details of the compactification, and hybrid networks of bound states can populate intermediate values~\cite{Copeland:2003bj,Jackson:2004zg}. Constraints on $P$, in conjunction with the dimensionless string tension $G\mu$, are therefore central to the observational programme on cosmic (super)strings.

Namikawa, Yamauchi and Taruya~\cite{Namikawa:2013wda} (hereafter NYT13) introduced the use of the curl-mode CMB lensing power spectrum as a probe of cosmic strings. Following earlier work on the lensing signal generated by vector perturbations~\cite{Yamauchi:2012bc,Yamauchi:2013fra}, they showed that, in the velocity-dependent one-scale (VOS) model~\cite{Martins:1995tg,Martins:2000cs,Avgoustidis:2005nv}, the curl-mode angular power spectrum induced by a string network scales approximately as $\Cww\propto (G\mu)^2 P^{-2}$ on the relevant large angular scales. As a consequence, $\Cww$ measurements primarily constrain the combination $G\mu P^{-1}$. From a curl-mode reconstruction of the ACT 2008-season temperature data, NYT13 obtained $G\mu \le 8.9\times 10^{-4}$ ($2\sigma$, $P=1$) and $G\mu P^{-1} \le 3.2\times 10^{-4}$ ($2\sigma$, small $P$); these bounds were limited primarily by the modest sky coverage and instrumental noise of the early ACT maps. NYT13 additionally derived tighter constraints from the Planck 2013 curl-mode reconstruction, $G\mu \le 6.6\times 10^{-5}$ ($2\sigma$, $P=1$) and $G\mu P^{-1} \le 3.4\times 10^{-5}$ ($2\sigma$, small $P$), which benefits from full sky coverage and access to lower curl multipoles down to $L_{\rm min}=2$. However, the lowest Planck multipole bins are known to be susceptible to mean-field and large-scale systematics, so that an independent constraint based on a ground-based reconstruction with a different scanning strategy, foreground treatment, and systematics budget is highly valuable as a cross-check of any cosmic-string interpretation.

This paper uses the recently released ACT DR6 curl-mode bandpowers, which supersede the ACT 2008 reconstruction by a wide margin~\cite{Qu:2023jia,Madhavacheril:2023vbh,MacCrann:2023zrh}. Recent analyses of the ACT DR6 primary CMB anisotropies have also yielded the stringent constraint $G\mu<3.66\times10^{-8}$ at $2\sigma$ under the assumption $P=1$~\cite{Raidal:2026cpb}. Compared with the ACT 2008 analysis, the survey area is larger by an order of magnitude, the noise level is substantially lower, and a profile-hardened, cross-correlation-based quadratic estimator is used to strongly suppress non-Gaussian foreground contamination. The signal-to-noise ratio of the lensing power spectrum is comparable to that of Planck that allows us to cross-check the Planck lensing measurement using ACT DR6 data. 
The DR6 curl reconstruction is found to be consistent with zero with a probability-to-exceed (PTE) of $0.37$ in the minimum-variance combination, providing both a stringent null test of the lensing pipeline and a competitive constraint on cosmological sources of curl power. 

In this paper we apply the NYT13 likelihood framework to the ACT DR6 curl-mode bandpowers to obtain new constraints on $G\mu$ and $P$. We work throughout in units where the curl-mode auto-spectrum is denoted $\Cww$, in line with the ACT DR6 convention and consistent across all equations in the present manuscript. Looking ahead, the Simons Observatory (SO)~\cite{SimonsObservatory:2018koc,SimonsObservatory:2025enh} will deliver a CMB lensing reconstruction with substantially improved sensitivity and access to lower lensing multipoles, both of which directly determine the achievable bound on the string parameters. The space-borne LiteBIRD satellite~\cite{LiteBIRD:2023iiy} will provide complementary leverage: as a full-sky polarisation experiment it can access the very lowest curl multipoles, and the polarisation-only reconstruction projected for the updated LiteBIRD configuration achieves a noise level comparable to, or better than, that of SO on the relevant large angular scales, particularly when combined with Planck~\cite{LiteBIRD:2025trg}. Further into the future, CMB-HD~\cite{CMB-HD:2022bsz} is expected to push the curl-mode noise floor substantially below current capabilities over half the sky. The framework developed here therefore provides the natural starting point for the upcoming SO analysis. Beyond the cosmic-string application studied here, the methodology developed in this paper is directly applicable to any source of vector or tensor metric perturbations, including primordial gravitational waves \cite{Dodelson:2003bt, Li:2006ss}, second-order scalar perturbations \cite{Cooray:2002mj,Saga:2015:second-order-curl, Pratten:2016dsm}, and exotic cosmological backgrounds 
such as Non-topological textures~\cite{Saga:2016yjd}.

Beyond the curl-mode reconstruction considered here, defect networks coupled to a pseudoscalar through the Chern-Simons interaction $(\phi/4 f_a)F\tilde F$ also generate a parity-violating rotation of the CMB linear polarisation~\cite{Jain:2021shf}. The sky-averaged (isotropic) component of this rotation is degenerate with the absolute polarisation-angle calibration of the instrument and is therefore set by the calibration floor, whereas the anisotropic component is not, since it is recovered via a quadratic estimator that measures the variance of the rotation field rather than its mean. The anisotropic-birefringence search at SO therefore provides a calibration-independent probe of the same defect networks, complementary to the curl-mode lensing route used here.

The remainder of the paper is organised as follows. Section~\ref{sec:theory} reviews the curl-mode angular power spectrum induced by a cosmic-string network in the VOS approximation. Section~\ref{sec:likelihood} describes the ACT DR6 curl-mode reconstruction and our likelihood. Section~\ref{sec:discussion} presents the resulting constraints on $G\mu$ and $P$, including a quantitative comparison to NYT13, and discusses the prospects for forthcoming SO data. Throughout the paper we adopt a flat \LCDM\ fiducial cosmology consistent with the analysis of Ref~\cite{Qu:2023jia}.

%% file: sections/2theory.tex
\section{Curl-mode lensing from a cosmic-string network}\label{sec:theory}

\subsection{Curl-mode angular power spectrum from vector perturbations}

A cosmic-string network sources scalar, vector, and tensor metric perturbations at all redshifts between recombination and today. CMB photons traversing this network acquire a deflection angle whose curl component is sensitive to the vector and tensor sectors~\cite{Yamauchi:2012bc,Yamauchi:2013fra,Namikawa:2011cs}. On the angular scales of interest for the current analysis the tensor contribution to $\Cww$ is negligible, and the dominant signal arises from vector metric perturbations~\cite{Yamauchi:2013fra}. The corresponding angular power spectrum can be written as~\cite{Yamauchi:2012bc,Yamauchi:2013fra}
\begin{equation}\label{eq:Cww_master}
C_L^{\Omega\Omega} = 4\pi \frac{(L-1)!}{(L+1)!} 
\int_0^{\infty} \frac{\mathrm{d}k}{k} 
\left[ \int_0^{\chi_*} \frac{\mathrm{d}\chi}{\chi} \, 
\Delta_1(k,\chi) \, j_L(k\chi) \right]^2,
\end{equation}
where $\chi_*$ is the comoving distance to the last-scattering surface, $j_L$ is the spherical Bessel function of order $L$, and $\Delta_1(k,\chi)$ is the dimensionless equal-time auto power spectrum of the vector metric perturbation. As in NYT13, we adopt the factorisable ansatz in which the unequal-time correlator is approximated by the geometric mean of the equal-time spectra at the two times.

\subsection{Velocity-dependent one-scale model}

To evaluate $\Delta_1^2(k,\chi)$ we adopt the VOS model of Martins and Shellard~\cite{Martins:1995tg,Martins:2000cs}, in which the network is characterised by a correlation length $\xi=1/(H\gamma)$ and a root-mean-square coherent velocity $v$, with $H$ denoting the Hubble expansion rate. In the radiation- and matter-dominated regimes, the network attains a scaling solution in which $\xi$ tracks the horizon. For small reconnection probabilities ($\tilde{c}P\ll 1$) the scaling values of $\gamma$ and $v$ admit the analytic approximations~\cite{Martins:2000cs,Avgoustidis:2005nv}
\begin{align}
\gamma &\simeq \left(\frac{\pi\sqrt{2}}{3\,\tilde{c}\,P}\right)^{1/2},
\label{eq:vos_gamma}\\
v^{2} &\simeq \frac{1}{2}\left(1 - \frac{\pi}{3\gamma}\right),
\label{eq:vos_v}
\end{align}
where $\tilde{c}\simeq 0.23$ encodes the efficiency of energy loss into loop production~\cite{Martins:2000cs}. The dimensionless vector power spectrum then takes the form~\cite{Yamauchi:2013fra}:

\begin{equation}
\begin{aligned}
\Delta_1^{2}(k,\chi)
={}&(16\pi G\mu)^{2}
\frac{\sqrt{6\pi}\,v^{2}}{12(1-v^{2})}
\frac{4\pi k^{3}\chi^{2}a^{4}}{H}
\\
&\times
\left(\frac{a}{k\xi}\right)^5
\operatorname{erf}\!\left(
\frac{k\xi/a}{2\sqrt{6}}
\right).
\end{aligned}
\label{eq:Delta1}
\end{equation}
where $a(\chi)$ is the scale factor.

\subsection{Parameter dependence and large-scale scaling}

Equation~\eqref{eq:Delta1} makes manifest the parameter scaling that drives the constraint geometry of curl-mode searches. 
Isolating the parametric dependence one finds, on the large angular scales where the signal is dominant ($L\to 0$),
\begin{equation}
\Cww \;\propto\; (G\mu)^{2}\,P^{-2}.
\label{eq:scaling}
\end{equation}
The combination $G\mu P^{-1}$ is therefore the principal direction of sensitivity, while the orthogonal direction enters only through the higher-multipole shape of the spectrum. As stressed in NYT13, this near-degeneracy implies that constraints in the $(P,G\mu)$ plane appear as bands of nearly fixed slope on a logarithmic plot. We confirm this behaviour explicitly in Section~\ref{sec:discussion}.

A second important consequence 
is that, for the parameter values of cosmological interest, the string-induced curl power spectrum is strongly red, peaking at the lowest accessible multipoles. The achievable constraint on $G\mu P^{-1}$ is therefore set primarily by the lowest multipole $\Lmin$ of the curl bandpower estimate and by its associated noise. This observation is the central rationale for re-applying the NYT13 framework to ACT~DR6 data, which provides the lowest-noise and best-controlled large-scale ($L\gtrsim 10$) curl reconstruction available from a single ground-based experiment.

%% file: sections/3likelihood.tex
\section{Data and likelihood}\label{sec:likelihood}

\subsection{ACT DR6 curl-mode bandpowers}\label{sec:dr6_curl}

We work with the curl-mode lensing reconstruction released as part of the ACT DR6 lensing analysis~\cite{Qu:2023jia}. The DR6 maps comprise five seasons of ACT observations (2017--2021) at frequencies centred on $90$ and $150$ GHz over $9400~\mathrm{deg}^{2}$ of sky, with a coadded white-noise level reaching ${\sim}10\,\mu\text{K-arcmin}$ in temperature 
(${\sim}14\,\mu\text{K-arcmin}$ in polarisation) in the deepest regions. The lensing 
analysis uses CMB multipoles in the range $600 \leq \ell \leq 3000$ for both temperature 
and polarisation, after masking point sources and a Galactic cut.

The lensing deflection field $\hat{\bm d}$ is reconstructed via the curved-sky quadratic estimator (QE) of Okamoto and Hu~\cite{Okamoto:2003zw}, in the implementation described in~\cite{Qu:2023jia}. The deflection field is decomposed into gradient ($\phi$) and curl ($\Omega$) components by acting with $\pm 1$ spin-weighted spherical harmonic transforms on the spin-1 displacement,
\begin{equation}
{}_{\pm 1}\hat{\bm d}(\hat{\bm n}) \;=\; \mp \!\sum_{LM}\,\frac{\bar\phi_{LM}\pm i\bar\Omega_{LM}}{\sqrt{L(L+1)}}\;{}_{\pm 1}Y_{LM}(\hat{\bm n}),
\label{eq:spin1_displacement}
\end{equation}
where overbars denote unnormalised quantities. Two key features of the DR6 pipeline are particularly relevant for the present analysis:

\begin{itemize}
\item \emph{Profile-hardened estimator.} Following the source-hardened construction of~\cite{Namikawa:2012pe} and the extragalactic-foreground-targeted variant of~\cite{Sailer:2020lal,Qu:2023jia}, the QE is bias-hardened against the leading non-Gaussian contamination from extragalactic point sources and the thermal Sunyaev--Zel'dovich (tSZ) effect. This significantly suppresses systematic foreground biases relative to the standard QE.
\item \emph{Cross-split-based estimator.} Rather than computing the auto-spectrum of a single reconstruction map, the DR6 pipeline cross-correlates lensing reconstructions obtained from disjoint splits of the data~\cite{Madhavacheril:2020eys,Qu:2023jia}. This eliminates the realisation-dependent Gaussian (\(N^{(0)}_L\)) bias arising from instrumental noise to a high degree, and removes the dependence of the recovered bandpowers on the detailed noise model.
\end{itemize}

The mean-field arising from masking and inhomogeneous noise is subtracted using the average reconstruction over $180$ noiseless full-sky simulations, while the residual lensing biases ($N^{(0)}_L$ from gaussian disconnected part and the lensing-dependent $N^{(1)}_L$ \cite{Kesden:2003jd, Hanson:2010rp}) are computed using 400 end-to-end Monte Carlo simulations \cite{Qu:2023jia}. A multiplicative Monte Carlo correction $\Delta A^{\rm MC,mul}_L$ is applied to the bandpowers to absorb residual normalisation effects from Fourier-space filtering and sky masking; this correction is at the few-percent level on the scales of interest.

For the curl-mode null test, the data products and analysis choices are identical to the baseline lensing analysis except that the spin-$\pm 1$ transforms in Eq.~\eqref{eq:spin1_displacement} are projected onto the $\Omega_{LM}$ component. The minimum-variance (MV) coadd combining temperature- and polarisation-derived estimators yields the curl-mode bandpowers $\widehat{C}_b^{\Omega\Omega}$ used in this work. As reported in~\cite{Qu:2023jia}, the MV curl bandpowers are consistent with zero with PTE $=0.37$, and the temperature-only (TT) variant has PTE $=0.75$. We adopt the MV curl bandpowers as our baseline.

\subsection{Bandpower compression}

We use the released MV curl-mode bandpowers binned according to the ACT DR6 baseline binning~\cite{Qu:2023jia}, with $20$ logarithmically spaced bins covering $40\le L\le 2000$. The lowest accessible bin centre is $L_{b=1}\simeq 15$, which (as discussed in Section~\ref{sec:theory}) is the multipole regime that drives the constraint on $G\mu P^{-1}$. We retain all bins for the constraint analysis; restricting to $L<200$ does not measurably change the resulting bounds, consistent with the expectation that the highest-$L$ bandpowers carry essentially no information on the cosmic-string parameters.

\subsection{Likelihood and parameter inference}

We compare the measured bandpowers to the theoretical curl-mode spectrum predicted by Eqs.~\eqref{eq:Cww_master}--\eqref{eq:Delta1} using a Gaussian likelihood,
\begin{equation}
-2\,\ln\mathcal{L}(G\mu,P) \;=\; \sum_{b=1}^{N_{\rm bin}}\,\frac{\bigl[\widehat{C}_b^{\Omega\Omega} - C_b^{\Omega\Omega,\,\mathrm{theo}}(G\mu,P)\bigr]^{2}}{\bigl(\sigma_b^{\Omega}\bigr)^{2}},
\label{eq:likelihood}
\end{equation}
where $\widehat{C}_b^{\Omega\Omega}$ denotes the binned curl-mode bandpower from the DR6 MV reconstruction, $C_b^{\Omega\Omega,\,\mathrm{theo}}(G\mu,P)$ is the theoretical bandpower computed for given $(G\mu,P)$ using the same bandpower window functions, and $\sigma_b^{\Omega}$ is the per-bin standard deviation drawn from the diagonal of the DR6 curl covariance matrix. We have verified that off-diagonal elements of the bandpower covariance are sub-dominant on the multipole range driving the constraint, so that the diagonal approximation in Eq.~\eqref{eq:likelihood} is adequate at the precision of this analysis. We adopt flat priors $\log_{10}P\in[-8,0]$ and $\log_{10}G\mu\in[-10,-6]$ and sample the likelihood on a dense $(P,G\mu)$ grid; for the special case $P=1$ we marginalise to a one-dimensional posterior $p(G\mu)$.

We follow the parametrisation choices of NYT13 throughout, so that any difference between our results and those of~\cite{Namikawa:2013wda} can be attributed unambiguously to the change in the input bandpower data and not to modelling choices. In particular, we use the same form of the VOS auto power spectrum, the same factorisable unequal-time ansatz, and the same Bessel-function projection in Eq.~\eqref{eq:Cww_master}.

\subsection{Combination with the Planck 2013 curl-mode bandpowers}\label{sec:planck13}

Because the cosmic-string curl-mode signal is strongly red and peaks at the lowest multipoles, the achievable bound on $G\mu P^{-1}$ is set predominantly by the minimum accessible curl multipole $\Lmin$. The publicly released curl-mode reconstructions therefore differ substantially in their constraining power: the ACT DR6 bandpowers reach $\Lmin=15$, whereas the Planck 2018 lensing release reports its curl-mode bandpowers only from $\Lmin=18$ (see Fig.~26 of Ref.~\cite{Planck:2018lbu}). By contrast, the Planck 2013 curl-mode reconstruction \cite{Planck:2013mth} used in NYT13 extends down to $\Lmin=2$, giving access to precisely the regime in which the string-induced signal is largest. For this reason we combine the ACT DR6 curl bandpowers with the Planck 2013 curl-mode bandpowers (rather than with Planck 2018) when forming a joint constraint. The two data sets are essentially independent on the relevant sky areas and at the lowest multipoles, and we add their contributions to the log-likelihood of Eq.~\eqref{eq:likelihood} assuming negligible cross-covariance. Throughout this paper our headline projection is the ACT DR6-only constraint, with the joint ACT DR6 + Planck 2013 result quoted alongside for comparison.

%% file: sections/4discussion.tex
\section{Results and Discussion}\label{sec:discussion}
\subsection{Constraints in the $(P,G\mu)$ plane}
\begin{figure}[t]
\centering
\includegraphics[width=\columnwidth]{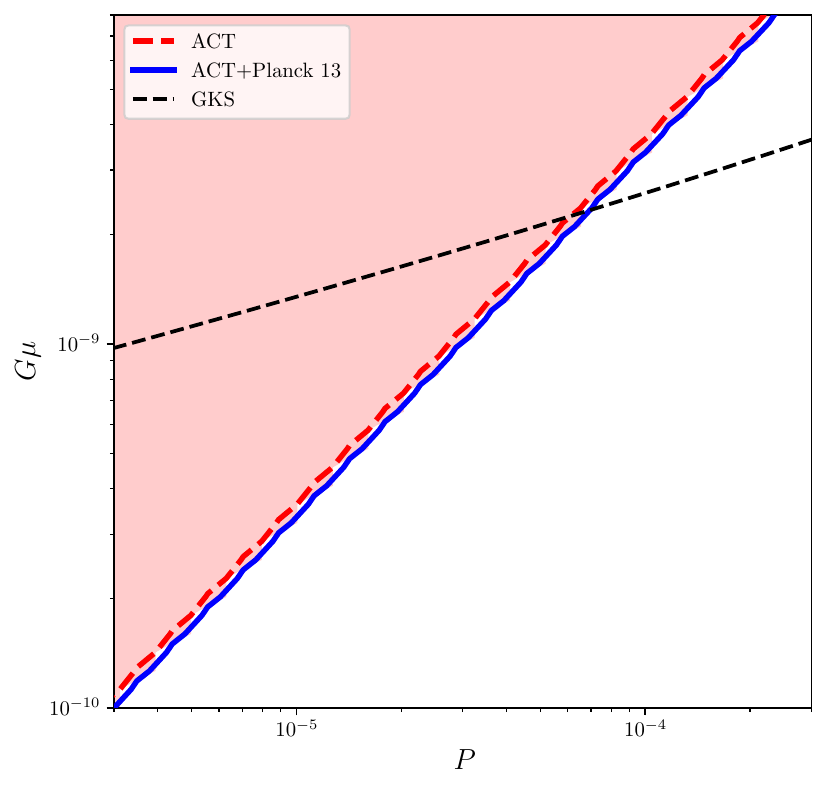}
\caption{Two-dimensional constraints on the cosmic-string parameters $(P,G\mu)$ from the ACT DR6 ($2\sigma$ upper limit, red dashed line) and joint ACT DR6 + Planck 2013 (blue solid line) curl-mode lensing bandpowers; the shaded region is excluded at $2\sigma$ by the joint constraint. The black dashed line shows the small-$P$ disfavoured region from the GKS-induced temperature angular power spectrum~\cite{Yamauchi:2010ms}.}
\label{fig:GmuP_constraint}
\end{figure}

Figure~\ref{fig:GmuP_constraint} shows the resulting two-dimensional likelihood contours in the $(P,G\mu)$ plane obtained from the ACT DR6 curl-mode bandpowers. For comparison we also show the joint constraint obtained by combining the ACT DR6 bandpowers with the Planck 2013 curl-mode reconstruction (see Section~\ref{sec:planck13}). As anticipated by the parameter scaling $\Cww\propto (G\mu)^{2}P^{-2}$ derived in Section~\ref{sec:theory}, the contours are very nearly straight lines of slope unity on the logarithmic plot, confirming that the measurement constrains the combination $G\mu P^{-1}$ rather than $G\mu$ and $P$ independently. Fitting the $2\sigma$ contour we find
\begin{equation}
G\mu P^{-1} \;\le\; 3.5\times 10^{-5}\qquad (2\sigma,\;P\lesssim 10^{-2}),
\label{eq:GmuP_bound}
\end{equation}
from the ACT DR6 bandpowers alone, tightening to
\begin{equation}
G\mu P^{-1} \;\le\; 3.2\times 10^{-5}\qquad (2\sigma,\;P\lesssim 10^{-2}),
\label{eq:GmuP_bound_joint}
\end{equation}
when the Planck 2013 curl-mode bandpowers are added. The two bounds are indicated by the red dashed (ACT only) and blue solid (ACT + Planck 2013) lines in Figure~\ref{fig:GmuP_constraint}, with the shaded region marking the parameter space excluded at $2\sigma$ by the joint constraint. For comparison we also overlay the small-$P$ disfavoured region from the temperature angular power spectrum via the Gott--Kaiser--Stebbins (\GKS) effect~\cite{Yamauchi:2010ms} as a black dashed line. As in NYT13, the curl-mode bound supersedes the GKS bound at sufficiently small reconnection probability; the tighter constraint obtained here pushes this transition to lower $P$ than in NYT13, so that at small $P$ the curl-mode constraint from ACT DR6 (and the joint ACT DR6 + Planck 2013 combination) is the dominant probe of the cosmic-string parameter space over a wider range of reconnection probabilities than was previously the case. 

We find that the ACT DR6 constraint is close to the Planck 2013 result. This indicates consistency between the ACT DR6 constraint and the previous Planck constraint obtained using multipoles down to ($L_{\min}=2$). The agreement suggests that the constraints are robust against possible systematic effects in the lowest multipole bins.

\subsection{Posterior on $G\mu$ at $P=1$}

\begin{figure}[t]
\centering
\includegraphics[width=\columnwidth]{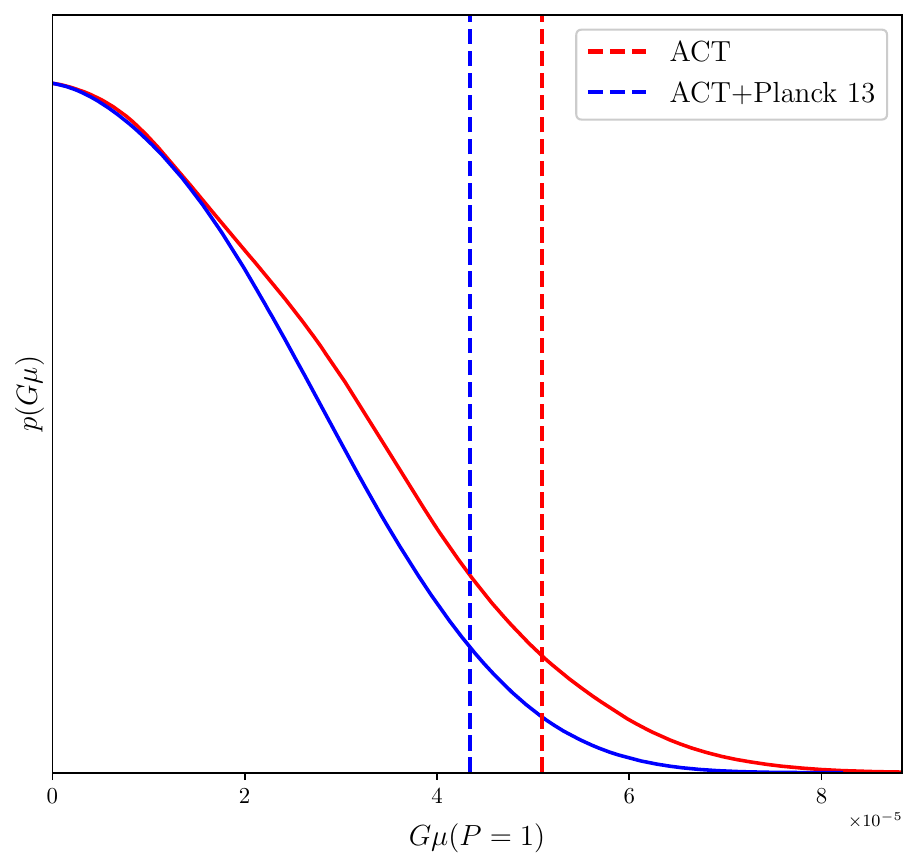}
\caption{Marginalised one-dimensional posterior $p(G\mu)$ at fixed $P=1$, obtained from the ACT DR6 curl-mode bandpowers (red) and from the joint ACT DR6 + Planck 2013 likelihood (blue). The dashed vertical lines indicate the corresponding $2\sigma$ upper bounds, $G\mu\le 5.0\times 10^{-5}$ (ACT) and $G\mu\le 4.3\times 10^{-5}$ (ACT + Planck 2013).}
\label{fig:Gmu_posterior}
\end{figure}

For ordinary field-theoretic strings, the canonical Nambu--Goto value $P=1$ is appropriate. Marginalising the likelihood at fixed $P=1$ yields the one-dimensional posterior $p(G\mu)$ shown in Figure~\ref{fig:Gmu_posterior}. The posterior peaks at $G\mu=0$, consistent with the curl-mode bandpowers themselves being statistically consistent with zero, and the corresponding $2\sigma$ upper bound is
\begin{equation}
G\mu \;\le\; 5.0\times 10^{-5}\qquad (2\sigma,\;P=1).
\label{eq:Gmu_bound}
\end{equation}
Including the Planck 2013 curl-mode bandpowers in the joint likelihood tightens this to
\begin{equation}
G\mu \;\le\; 4.3\times 10^{-5}\qquad (2\sigma,\;P=1).
\label{eq:Gmu_bound_joint}
\end{equation}

\begin{figure}[t]
\centering
\includegraphics[width=\columnwidth]{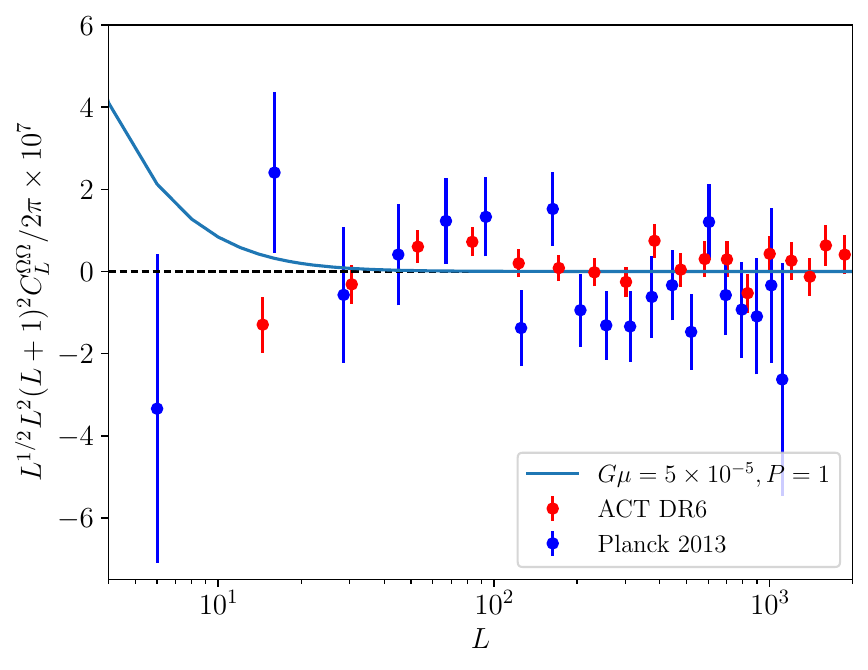}
\caption{ACT DR6 minimum-variance curl-mode bandpowers (red points) and Planck 2013 curl-mode bandpowers (blue points) compared with the theoretical curl-mode angular power spectrum $\Cww$ predicted for cosmic strings with $G\mu=5\times 10^{-5}$ and $P=1$ (solid line), close to the $2\sigma$ upper bound from the ACT-only analysis (Figure~\ref{fig:Gmu_posterior}). The bandpowers are plotted with the scaling $L^{1/2}L^{2}(L+1)^{2}\Cww/(2\pi)$ following the convention of~\cite{Qu:2023jia}.}
\label{fig:Cl_comparison}
\end{figure}

To visualise the data-theory comparison underlying these bounds, in Figure~\ref{fig:Cl_comparison} we plot the ACT DR6 MV curl-mode bandpowers $\Cww$ together with the Planck 2013 curl-mode bandpowers, and overlay the theoretical curl-mode spectrum at $G\mu=5\times 10^{-5}$, $P=1$, close to the $2\sigma$ upper limit derived from Figure~\ref{fig:Gmu_posterior} for the ACT-only analysis. The string-induced curl power spectrum is strongly localised at the lowest multipoles, with virtually no signal above $L\sim 50$. The constraint is therefore set almost entirely by the lowest two ACT DR6 bins, which span $L\sim 10$--$40$, together with the $L<15$ bins of Planck 2013 in the joint analysis.

\subsection{Constraints with conservative multipole ranges}\label{sec:lge40}

The ACT DR6 lensing analysis~\cite{Qu:2023jia} reports that, for $L > 40$, the distribution of null-test PTEs across both map-level and bandpower-level tests is consistent with a uniform distribution and shows no significant evidence of systematic contamination. Extending to lower multipoles, the distribution exhibits a few borderline failures; the principal concern at low $L$ is the size of the mean-field bias in the gradient reconstruction, which motivates the $L_{\min}=40$ cut for cosmological interpretation. Since the mean-field is smaller for the curl, the scale cut motivated by the gradient mean-field would not specifically apply to it. For the Planck 2013 data, the conservative multipole range for the lensing potential is ($40 \leq L \leq 400$) \cite{Planck:2013mth}, as the lower multipole bins are affected by systematic uncertainties associated with the mean-field bias and foregrounds. Moreover, the curl-mode power spectrum was not extensively validated, leaving the robust multipole range unclear.

As a conservative cross-check, we therefore repeat the analysis restricting both the ACT DR6 and Planck 2013 bandpowers to $L>40$. We find $G\mu \leq 4.1\times10^{-4}$ ($2\sigma$, $P=1$) from ACT DR6 alone and $G\mu \leq 4.05\times10^{-4}$ ($2\sigma$, $P=1$) when combining ACT DR6 with the Planck 2013 curl-mode bandpowers. The substantial degradation when discarding the low-$L$ bins quantifies the central role played by $\Lmin$ and motivates the use of the full multipole range as the baseline. The dependence of the constraints on $\Lmin$ is summarized in Table~\ref{tab:lge40}.

\begin{table}[t]
\caption{$2\sigma$ upper bounds on $G\mu$ at $P=1$ for different minimum multipoles applied to both the ACT DR6 and Planck 2013 curl-mode bandpowers.}
\label{tab:lge40}
\begin{ruledtabular}
\begin{tabular}{ccc}
$\Lmin$ & ACT DR6 & ACT DR6 + Planck 2013 \\
\hline
2  & $5.04\times10^{-5}$  & $4.31\times10^{-5}$  \\
20 & $1.67\times10^{-4}$ & $1.64\times10^{-4}$ \\
40 & $4.13\times10^{-4}$  & $4.05\times10^{-4}$ \\
\end{tabular}
\end{ruledtabular}
\end{table}

\subsection{Comparison with previous curl-mode constraints}
Table~\ref{tab:comparison} summarises the present constraints alongside those obtained by NYT13 from the ACT 2008-season curl reconstruction. Two complementary measures of improvement are quoted: the absolute $2\sigma$ upper bound on $G\mu$ for $P=1$, and the small-$P$ bound on the combination $G\mu P^{-1}$.

The pattern of improvements reflects the underlying experimental drivers. The constraint on $G\mu$ for $P=1$ is set by the moderate-$L$ behaviour of the curl bandpowers, where the much lower noise floor of the DR6 reconstruction yields a more than order-of-magnitude improvement over the ACT 2008 result. The bound on the small-$P$ combination $G\mu P^{-1}$, by contrast, is set primarily by the lowest accessible curl multipoles. Both the DR6 and the ACT 2008 binning of NYT13 reach down to $L\sim 10$, so the gain in this regime is driven entirely by the substantially reduced variance per bin, yielding the approximately $90\%$ (factor-of-ten) improvement quoted in Table~\ref{tab:comparison}. 

It is worth emphasising that the ACT DR6 curl-mode reconstruction has a number of qualitative advantages over the curl-mode bandpower estimate used in NYT13. The profile-hardened cross-correlation-based estimator~\cite{Qu:2023jia} is much more robust to extragalactic foreground non-Gaussianity than the bias-hardened estimator employed in NYT13, and the cross-correlation construction eliminates noise mean-field contamination that affected the standard quadratic estimator used at the time. The ACT DR6 curl bandpowers also pass a comprehensive suite of map-level null tests~\cite{Qu:2023jia}. The bounds quoted in Table~\ref{tab:comparison} are therefore not only quantitatively but also qualitatively more robust than their NYT13 counterparts.

\begin{table}[h]
\caption{Comparison of $2\sigma$ upper bounds on the cosmic-string parameters from CMB curl-mode lensing reconstructions of ACT data. The ACT 2008-season values are from NYT13; the ACT DR6 
values are from the present analysis based on the curl reconstruction of~\cite{Qu:2023jia}. The right-most column gives the relative improvement of the present ACT DR6 work over the ACT 2008 bound.}
\label{tab:comparison}
\begin{ruledtabular}
\begin{tabular}{lccc}
 Parameter & ACT 2008 & ACT DR6 & Improvement\\
\hline
$G\mu\,(P=1)$ & $8.9\times 10^{-4}$ & $5.0\times 10^{-5}$ & $94.4\%$\\
$G\mu P^{-1}$ & $3.2\times 10^{-4}$ & $3.5\times 10^{-5}$ & $89.1\%$\\
\end{tabular}
\end{ruledtabular}
\end{table}

%% file: sections/5conclusions.tex
\section{Conclusions}\label{sec:conclusions}
We have applied the curl-mode cosmic-string framework of NYT13 to the ACT DR6 curl-mode lensing bandpowers and obtained new $2\sigma$ upper bounds on the dimensionless string tension and the reconnection probability,
\begin{align}
    G\mu P^{-1} &\;\le\; 3.5\times 10^{-5} \quad (P\lesssim 10^{-2})
    \,, \\
    G\mu &\;\le\; 5.0\times 10^{-5} \quad (P=1)
    \,. 
\end{align}
The consistency between the ACT DR6 and Planck 2013 constraints suggests that the results are robust against possible systematic effects in the lowest multipole bins.
Combining ACT DR6 with the Planck 2013 curl-mode bandpowers, which alone reach down to $\Lmin=2$, tightens these to 
\begin{align}
    G\mu P^{-1} &\;\le\; 3.2\times 10^{-5} \quad (P\lesssim 10^{-2})
    \,, \\
    G\mu &\;\le\; 4.3\times 10^{-5} \quad (P=1)
    \,. 
\end{align}
To our knowledge these are the tightest constraints on cosmic strings derived from a CMB curl-mode lensing measurement reported to date. 
Focusing on the ACT data alone, the bound on ($G\mu P^{-1}$) from ACT DR6 improves upon that from the ACT 2008 season by nearly an order of magnitude, while providing significantly greater robustness against systematics owing to the profile-hardened, cross-split-based pipeline.
Additional data from ACT, SO, LiteBIRD, and CMB-HD can further improve these constraints.

%% file: sections/6akg.tex
\begin{acknowledgments}
We thank Colin Hill, Arthur Kosowsky, Alexander van Engelen, Kam Arnold and Sebastian Belkner for helpful comments and the ACT collaboration for making the DR6 curl-mode bandpowers publicly available. We also thank the Planck team for kindly providing us with the curl-mode power spectrum obtained from the Planck temperature data. We acknowledge the use of the public CAMB~\cite{Lewis:1999bs}, NumPy~\cite{Harris:2020}, SciPy~\cite{Virtanen:2020}, emcee~\cite{ForemanMackey:2013}, GetDist~\cite{Lewis:2025}, and Matplotlib~\cite{Hunter:2007} software packages. TN acknowledges support from JSPS KAKENHI Grant No. JP25K00996, and No. JP26H00405, and from JST EXPERT-J, Japan Grant No. JPMJEX2508. The Kavli IPMU is supported by World Premier International Research Center Initiative (WPI Initiative), MEXT, Japan. 
\end{acknowledgments}

%% file: sections/7data.tex
\section*{Data and Code availability}
The ACT DR6 lensing data products used in this analysis, including the curl-mode bandpowers, bandpower window functions, and the bandpower covariance matrix, are publicly available through the LAMBDA archive\footnote{\url{https://lambda.gsfc.nasa.gov/}} and the ACT collaboration data release page, as described in~\cite{Qu:2023jia}. The theoretical curl-mode template generation code based on the velocity-dependent one-scale model is available at \url{https://github.com/antolonappan/cosmic_strings}.